# Bioinspired multi-asymmetric magnetized surfaces for tailored liquid operations and 3-DOF solid transport

Jiaqi Miao[1]*

[1]Department of Mechanical Engineering, The University of Hong Kong, Pokfulam, Hong Kong, China
(Email: jqmiao@connect.hku.hk)

**Abstract:** Through the utilization of smart materials and well-designed structures, functional surfaces have been developed to enable small-scale liquid/solid manipulation tasks, thereby facilitating crucial applications in the fields of microfluidics, soft robotics, and biomedical engineering. However, the design of functional systems with flexible, tunable, and multimodal liquid/solid manipulation capabilities remains a challenging endeavor. Here, inspired by asymmetric structural features in natural plants and metachrony in cross-scale biological systems, I report a magnetic-responsive functional surface that can achieve rich liquid operations under static magnetic fields, while also enabling the transportation of solids with multiple degrees of freedom (DOFs) under dynamic magnetic fields. The presence of curvature pillars on the surface, combined with their magnetic-driven tilt/gradient arrangement, imparts liquids with multi-directional spreading modes based on the asymmetry of Laplace pressure. I elucidate the mechanisms governing these liquid spreading modes and subsequently develop compelling liquid operations, such as adjustable anti-gravity climbing, spontaneous modal shifts in liquid transport, and liquid mixing. Furthermore, the dynamic metachronal motion of the magnetic pillars can be harnessed for solid object transportation. I illustrate the synchronous/asynchronous transport modes of the surface and propose a novel strategy for achieving 3-DOF solid transportation by coordinating the arrangement of objects and employing magnetic actuation strategies. This study presents a new design concept for application-oriented manipulation surfaces, which hold significant potential for extensive engineering applications.

# 1. Introduction

Small-scale liquid and/or solid transport is ubiquitous in nature, ranging from static plants that can directionally guide liquid flow to motile organisms that can actively interact with environments [1-5]. These transport behaviors support extensive biological activities, including feeding, locomotion, environmental exploration, and sensing. Elucidating the intrinsic mechanisms underlying these behaviors not only represents significant progress in the life sciences but also presents valuable opportunities in engineering. Enlightened by the remarkable transport capabilities observed in nature, researchers have developed a myriad of biomimetic functional systems to manipulate liquids and solids. Broadly, these systems can be classified into two categories based on their operating modes: static and dynamic. Structured surfaces, serving as prominent examples of static systems, have been proposed for stable and energy-efficient liquid transport [6-11]. These functional surfaces, featuring intricate structures inspired by natural plants such as Nepenthes [1], cactus [12], spider silk [13], and Araucaria [14], can facilitate unidirectional liquid spreading through the asymmetry of physical structures and surface energy. On the other hand, dynamic systems employ more intuitive approaches to small-scale manipulation, e.g., arrays of micro-pillars/needles for liquid/solid transportation [15-20], biomimetic grippers [21-24], and soft actuators [25-28]. By utilizing stimuli-responsive materials that respond to light, heat, electricity, magnetic fields, or pH changes, these small-scale dynamic systems can accomplish a diverse range of manipulation tasks with high flexibility. These collective efforts have significantly contributed to various engineering applications, involving water harvesting [29], thermal management [30,31], microfluidics [32-35], small-scale manipulation [36-38], soft robotics [39,40], and biomedical applications [41,42].

However, both static and dynamic liquid/solid manipulation systems for liquid/solid manipulation encounter several challenges. In the case of liquid operations, static systems' manipulation flexibility is greatly limited by their fixed structures, while dynamic systems need complex external equipment and precision control [43,44]. Concerning solid transportation, although researchers have developed responsive surfaces capable of transporting solid objects with 1-DOF modes (such as back and forth motion [38], or unidirectional transport [45]) and even 2-DOF modes enabling arbitrary in-plane directions [18,44,46,47]); the exploration of small-scale surfaces with increased transport degrees of freedom remains unexplored. Consequently, biomimetic systems for small-scale liquid and/or solid manipulation are still in their infancy, whether it be for striking a balance between fixed and responsive surfaces or enhancing the flexibility of solid transportation (e.g., by incorporating more DOFs).

To augment manipulation flexibility and functionality, a promising avenue is the design of responsive surfaces informed by surface science principles [15,48-50]. For instance, Jiang et al. used the modified superhydrophobic surface to directionally transport droplets, where droplets are captured via surface adhesion and released under magnetic control [51]. Furthermore, the application of magnetic fields has recently been shown to enable the reconfiguration of surfaces, altering the direction of liquid spreading [49] or enhancing liquid regulation [52]. Unique surface designs have also been employed to enable unidirectional solid transport while impeding transport in the opposite direction [45]. These examples, combining multiple biomimetic designs with the responsive ability to external stimuli, offer promising prospects for manipulating liquids and solids. However, it is imperative to engage in further exploration and development to effectively address the existing challenges associated with transport mechanisms, magnetic control methodologies, and novel manipulation modes.

Herein, I present a novel bioinspired multi-asymmetric magnetized surface capable of facilitating diverse liquid operations and 3-DOF solid transport (Fig. 1a). Drawing inspiration from the asymmetrical characteristics of natural plants like Nepenthes, ryegrass leaf, and cactus spine, I aim to achieve a comprehensive system that incorporates curvature, tilt, and gradient configurations (Fig. 1b). Initially, the surface's curvature pillars generate asymmetric Laplace pressure, directing the controlled spreading of liquids (Fig. 1b(i)). Subsequently, by applying external magnetic fields, the magnetic-responsive pillars reconfigure into tilted or gradient arrangements (Fig. 1b(ii)-(iii)). Notably, in conjunction with the curvature pillars, these new configurations enable various liquid transport modes, characterized by enhanced speed and altered direction (Fig. 1b(iv)). I reveal the underlying mechanisms and develop innovative liquid operations such as adjustable anti-gravity transport, spontaneous modal shift in liquid transport, and liquid mixing. Additionally, by replicating the metachrony in cross-scale biological systems (Fig. 1c(i)-(iii)), I demonstrate that pillars' dynamic metachronal coordination can be utilized to transport solid objects (Fig. 1c(iv)). Biologically, metachrony refers to sequential cyclic motion of appendages with a fixed phase difference. In this study, objects with different shapes and arrangements can be synchronously or asynchronously transported across the surface. Further, through coordination of object placement and magnetic control, I showcase the surface's ability to achieve novel 3-DOF solid transportation encompassing two translational and one rotational degrees of freedom. Overall, this study presents a new design concept for application-oriented liquid/solid manipulation surfaces.

## 2. Experimental Section

### 2.1. Materials

Ecoflex 00-30 was purchased from Smooth-On, Inc. Iron microparticles with an average diameter of 25-48 μm and the ferrite magnet were obtained from the Chinese store. Ethanol (AR, water ≤ 0.3%) was provided by Aladdin Industrial Co. Ltd. (Shanghai, China). The used dyes include violet biodye (BKMAMLAB, China), pink biodye (BKMAMLAB, China), red ink and blue ink purchased from the Chinese store.

### 2.2. Fabrication of the magnetized surface

The surface was prepared using a molding process with the magnetically assisted arrangement. As depicted in Fig. 2a, the surface is composed of a 40 × 16 array of semicolumnar pillars. Each pillar has a height of 4 mm, and a cross-sectional semicircle diameter is 2 mm, the horizontal and vertical spacing between pillars are 1.6 mm and 3 mm, respectively. Abovementioned dimensions correspond to the motherboard mold, and the actual measured dimensions can be found in Table. S1. Fig. 2b shows the whole fabrication process of the magnetized surface. As shown in Fig. 2b(i), a 3D-printed mold is manufactured via the stereolithography (SLA) 3D printing technology (resolution: ± 0.1 mm; provided by the WENEXT company from Shenzhen, China; the used resin: Future 8200Pro Resin). The main material for the surface was Ecoflex 00-30, a platinum-catalyzed silicone elastomer formed by mixing Ecoflex A and Ecoflex B in a mass ratio of 1:1. To confer magnetic-responsive properties, iron microparticles were added to the soft material. For the molding process, a magnetic suspension, prepared by evenly mixing Ecoflex A, Ecoflex B, and iron microparticles (in a mass ratio of 1:1:1.5), was poured into the mold (Fig. 2b(ii)). After that, the mold was placed in the vacuum chamber to remove air bubbles and make the magnetic suspension thoroughly fill the mold (Fig. 2b(iii)). Subsequently, the upper layer of the magnetic suspension was carefully wiped off with new cotton swabs until it is completely clean. Then, a non-magnetic suspension, prepared by evenly mixing Ecoflex A and Ecoflex B in a mass ratio of 1:1, was poured into the mold as a replacement (Fig. 2b(iv)-(v)). During the curing process, the mold was positioned above a ferrite magnet measuring 150 mm × 100 mm × 25 mm, with the distance between the mold and magnet being almost zero (Fig. 2b(vi)). The larger size of the magnet, in comparison to the mold, ensured an approximate upward magnetic arrangement direction. A Gaussmeter (resolution: ±2%; Nohawk, Inc. from Tianjin, China) was used to measure the quasi-vertical magnetic field applied at this stage, which was approximately 0.12 T. The non-magnetic suspension effectively prevented the Rosensweig-type instability (caused by the magnetic field), ensuring the stability of solidification. Finally, the surface was obtained through a demolding process after curing for 8 hours at room temperature (25 °C).

## 2.3. Experimental setup for liquid operations and solid manipulation

The experimental setup for liquid operations consisted of the following components: a syringe pump (Longer Pump, LSP02-2B), a syringe connected to tubes and a needle, a ferrite magnet, a handheld microscope (AM4115ZTL Dino-lite Edge) used in conjunction with Dino Capture 2.0 software, and the surface. All data were processed using ImageJ software.

To achieve uniform bending and gradient arrangement for various liquid operation modes, an upright magnet (with the left pole as N and the right pole as S) was translated beneath the surface. The liquid injection rate was set at 500 μL/min. All used ethanol-water binary mixtures were made just before the experiments and immediately encapsulated into the syringe to minimize ethanol volatilization. The contact angle exhibited by the sealed binary mixture in the needle tube remained almost unchanged for half an hour (Fig. S1). Meanwhile, to confirm the uniformity of surface energy between the substrate material (pure Ecoflex 00-30) and the pillar material (Ecoflex 00-30 + iron microparticles), the contact angles of three different types of liquids (nearly 100% ethanol, DI water, and a binary mixture of 50% ethanol and 50% DI water) were measured on both materials. As shown in Fig. S2, the contact angles observed on the two surface materials exhibited similar values, indicating that they can be considered as the same material for theoretical analysis.

Additionally, dyes were employed to visualize experimental phenomena, and their influence on the surface tension of the liquids was evaluated. Since the dyes used were water-soluble, they were treated as part of the water component when preparing the binary mixture of DI water and ethanol. As illustrated in Fig. S3, the contact angles of all the unstained/stained liquids used in this study were measured on pure Ecoflex 00-30, and the results demonstrate that the inclusion of dyes has a negligible effect on the surface tension of the binary mixture of DI water and ethanol. Except for the dyed liquids, which were captured as color images (to visually see the color difference), the contact angle and surface tension measurements for the other liquids were obtained via an optical video-based contact angle meter (Model 100SB, Sindatek Instrument Co., Ltd.).

In the context of the solid transport experiments, the surface underwent metachronal motion achieved through manual translation manipulation of the magnet, with each cycle lasting approximately 4.5 seconds. Here, a cycle refers to the duration during which the surface is influenced by the magnetic field, spanning from the beginning to the end. The magnet actuates the surface from a distance of approximately 9 mm away. Taking into consideration the influence range of the magnetic field, the magnet's movement speed is estimated to be about 26 mm/s.

## 3. Results and Discussion

### 3.1. Calibration of the magnetic-driven tilt and gradient arrangement

As illustrated in Fig. 2c(i), the translated magnet, taking the translation from right to left as an example, can actuate the surface's pillars to present different arrangements, including the original standing (unaffected) and tilt/gradient arrangement (affected). Specifically, the magnetic field effect on the surface can be divided into five stages, which are characterized by their different arrangements. For easy identification, gray and orange regions are respectively used to represent standing pillars and uniform bending pillars; the gradient arrangement of pillars is classified as gradient (+) region in green (i.e., the gradient decrease direction is towards right) and gradient (-) region in blue (i.e., the gradient decrease direction is towards left). Meanwhile, some definitions are given to acquire intuitive quantitative data. When the magnet starts and stops affecting the surface, the position of the magnet's center axis is defined as the starting line and the terminal line, respectively. The length of the affected range, denoted as $s$, is closely associated with $h$, representing the distance between the bottom of the surface and the top of the magnet, assuming the surface and driving magnet remain the same. The distance between the central axis of the magnet and the starting line is defined as $L$ when the magnet moves from the starting line to the ending line. As $L$ increases from 0 to $s$, five influenced stages of the surface occur sequentially. During stage 1, the affected region is limited, exhibiting a gradient (+) state when stimulated by the gradient magnetic field distributed on the upper left side of the magnet. In stage 2, as the magnet is positioned below the surface, a quasi-uniform magnetic field above the magnet, pointing to the right, induces uniform bending of the local pillars. At this time, the affected regions consist of both the gradient (+) region and the uniformly bending region. In stage 3, the gradient magnetic field distributed on the upper right of the magnet begins to contribute to a gradient (-) region. For the next period of time, the gradient (+) region, uniform bending region, and gradient (-) region will exist simultaneously and move away from the starting line as the magnet translates. The region influenced by the magnetic field decreases in stages 4 and 5, causing the gradient (+) region and the uniform bending region to sequentially vanish. Once the magnet reaches the terminal line, the surface is no longer influenced by the magnetic field, leading to a complete recovery of its upright position, thereby signifying the conclusion of stage 5.

Fig. 2c(ii)-(iii) shows the quantitative results for the five stages. As the magnet is translated, the gradient (+) region, uniform bending region, and gradient (-) region sequentially emerge and progressively move away from the starting line. This is manifested by an increase in the distances from the central axis of each region to the starting line ($g_1$, $b$, $g_2$), as illustrated in Fig. 2c(ii). In the data graph, I have approximately divided the five stages, with each stage exhibiting distinct states of the pillars, including the standing state, uniform bending, and gradient (+)/(-) arrangement. Furthermore, Fig. 2c(iii)

illustrates that in stage 3 (where all three regions exist), the uniform bending region contains more affected pillars than the gradient (+)/(-) region. This outcome is attributed to the applied magnetic field. When the magnet is in closer proximity to the surface, the quasi-uniform magnetic field has a larger range than the gradient magnetic field at the edge. To provide further clarification, I assess how the surface is influenced by the strength and distribution of the magnetic field, which can be adjusted by modifying the parameter $h$.

In terms of the affected area, as expected, it will become smaller when the magnet is moved away from the surface. As shown in Fig. 2d, when $h$ changes from 6 mm to 18 mm, the length of the affected area ($s$) will change from approximately 127 mm to 82 mm. And predictably, the number of affected pillars and the magnetic effect (i.e., force and torque) acting on the pillars also decrease as the magnet is moved away. Here, the bending angle ($\theta$) of the pillars is defined as a measure of the magnetic effect. Fig. 2e illustrates a decreasing trend in the number of uniformly bent pillars, along with a continuous decrease in their bending angle $\theta$. It indicates a significant weakening of the quasi-uniform magnetic field above the magnet. To provide more precise quantitative information, the variation of magnetic field strength with $h$ is measured and presented in Fig. S4. Furthermore, these data illustrate the flexibility and tunability of the surface under different magnetic field strengths. Fig. S5 shows that the maximum bending angles observed in the two-side bending configuration, when $h = 0$, are about 50º. Larger bending angles correspond to smaller average curvature radii, and they also result in stronger Laplace pressure effects. These parameters that characterize the surface flexibility are influenced by various factors, including the strength of external magnetic fields, the dimensions of the surface, and the content of magnetic microparticles in pillars. During stage 3, as the value of $h$ increases, the uniform bending region exhibits a decrease in the number of affected pillars compared to the gradient region. Consequently, the proportion of uniform bending pillars to magnetic-affected pillars (in the bending region) diminishes, as depicted in Fig. 2f.

### 3.2. Mechanisms of liquid directional transport on the surface

To elucidate the underlying mechanisms of the directional liquid transport on the magnetized surface, I investigate the rectification effect resulting from the asymmetric features, including the intrinsic curvature structure, magnetic-field-induced bending, and gradient arrangement. Here, the ethanol-water binary mixture (15% ethanol + 85% water) is used to demonstrate the liquid spreading behavior. The continuous injection of the liquid through a syringe pump allows for a dynamically updated boundary condition in terms of volume flow. In this process, the semicolumnar pillars play a role in constructing unbalanced Laplace pressure, which further lead to the unidirectional spreading of the liquid. As shown in Fig. 3a(i), the liquid spreading on the surface can be considered as the flow in capillary tubes with varying diameters. By comparing the Laplace pressure on both sides, the continuously injected liquid will always move toward the direction with smaller pressure (spreading direction) and be pinned in the opposite direction with larger resistance (pinning direction). The Laplace pressure $\Delta p$ can be generally described as:

$$\Delta p = \gamma(1/r_1 + 1/r_2) \quad (1)$$

where $\gamma$ is the liquid surface tension, $r_1$ and $r_2$ are the two principal radii of the liquid front curvature. Here, we define the Laplace pressures on the spreading direction ($\Delta p_+$) and the pinning direction ($\Delta p_-$) by considering the liquid front curvature in the top view ($1/r_1$) and front view ($1/r_2$), respectively, which are expressed as:

$$\Delta p_+ = \Delta p_{1+} + \Delta p_{2+} = \gamma(1/r_{1+} + 1/r_{2+}) \quad (2)$$

$$\Delta p_- = \Delta p_{1-} + \Delta p_{2-} = \gamma(1/r_{1-} + 1/r_{2-}) \quad (3)$$

As shown in Fig. 3a(i), $d_+(x)$ and $\alpha(x)$ are the width and the widening angle of the capillary channel at the position $x$ under consideration. By using trigonometry, the derived capillary pressure $\Delta p_{1+}$ will be:

$$\Delta p_{1+} = -2\gamma \sin(\theta_a - \alpha(x))/d_+(x) \quad (4)$$

where $\theta_a$ is the advancing contact angle of the liquid. Similarly, we can derive the Laplace pressure $\Delta p_{1-}$ in the pinning direction. The critical condition in this direction is the disappearing capillary channel, which $\alpha(x)$ is equal to 0. Thus, we can get the equation as below:

$$\Delta p_{1-} = -2\gamma \sin(\theta_a)/d_-(x) \quad (5)$$

where $d_-(x)$ is the width of the capillary channel in the pinning direction. Further, we consider the liquid front curvature in the top view to acquire another component of Laplace pressure ($\Delta p_{2+}$ and $\Delta p_{2-}$). According to the illustration of Fig. 3a(ii), $\Delta p_{2+}$ and $\Delta p_{2-}$ are the same because they exhibit the same radius of curvature, which is described as:

$$\Delta p_{2+} = -\gamma(1 - \cos\theta_a)/h_+(x) = \Delta p_{2-} = -\gamma(1 - \cos\theta_a)/h_-(x) \quad (6)$$

where $h_+(x)$ and $h_-(x)$ represent the capillary channel height in the spreading direction and the pinning direction, respectively. Till now, we can express the total Laplace pressures $\Delta p_+$ and $\Delta p_-$ as:

$$\Delta p_+ = -2\gamma \sin(\theta_a - \alpha(x))/d_+(x) - \gamma(1 - \cos\theta_a)/h_+(x) \quad (7)$$

$$\Delta p_- = -2\gamma \sin(\theta_a)/d_-(x) - \gamma(1 - \cos\theta_a)/h_-(x) \quad (8)$$

Based on equation (7) and (8), it is obvious that $|\Delta p_+|$ is smaller than $|\Delta p_-|$, meaning that continuously injected liquids will always preferentially break the pinning in the spreading direction and present the unidirectional transport. As depicted in Fig. 3a(ii), because the unbalanced Laplace pressure causes the liquid to undertake relatively smaller resistance in the spreading direction, the liquid shows a smaller dynamic contact angle in the spreading direction, i.e., $\theta_{a+} < \theta_{a-}$ (where $\theta_{a+}$ and $\theta_{a-}$ are the dynamic contact angles in the spreading and pinning directions, respectively). Fig. 3a(iii) presents the unidirectional transport experiment of the liquid. Within 39 s, the spreading distance is about 17.6 mm (Video S1).

The magnetic field causes the uniform and gradient bending arrangement of pillars, enabling the reconfiguration of the surface into positive arrangement (used for accelerating the transport process) and negative arrangement (used for changing the spreading direction), as illustrated in Fig. 3b(i)-(ii). In the positive arrangement, the pillars exhibit a gradient (+) region on the left, followed by a region of uniform leftward bending, and finally a gradient (-) region (Fig. 3b(i)). When the liquid is introduced onto the gradient (-) region, it spreads continuously towards the left while being pinned on the right. This behavior is attributed to the leftward bending pillars generating a net Laplace pressure ($P_1$) on the left side, propelling the liquid forward. Simultaneously, the convex liquid surface on the right side exerts a Laplace pressure ($P_2$), preventing the liquid from spreading further. As the experimental snapshots show, the liquid first fills the gap between two pillars in the adjacent rows, then unidirectionally spreads towards left. In the negative arrangement of the surface, the pillars exhibit a gradient (+) region on the left, followed by a region of uniform rightward bending, and finally a gradient (-) region (Fig. 3b(ii)). The liquid is injected onto the gradient (+) region, and unidirectionally spreads towards right due to the pulling effect of the net Laplace pressure $P_4$, while pinned at the gradient (-) region because of the pinning effect of the net Laplace pressure $P_3$. As the liquid is continuously injected on the surface, it fills the gaps in the negative arrangement, then pinned by the curvature pillar structure on the standing region (without magnetic field effect).

Compared with the un-actuated surface, the positive arrangement enhances the speed of liquid transport, while the negative arrangement alters the direction of liquid spreading (Fig. 3c and Video S2). Specifically, after 24 seconds, the liquid spreads approximately 24 mm on the positive arrangement, whereas it only covers around 10 mm on the un-actuated surface. As for the negative arrangement, the liquid spreads approximately 19.5 mm in the opposite direction within 21 seconds, after which it becomes pinned. Such a pinning effect causes the liquid to accumulate and extend beyond the tip of the pillars (Fig. 3c and Video S3). It should be noted that even if the spreading behavior shows an approximate linear relationship, it may not fully imply a classical convective flow. Actually, the liquid spreading behavior entails a step-wise structural "filling" process [53]. The process of breaking the pinning of each row of pillars requires additional time, resulting in a non-linear speed for the overall spreading process.

Therefore, the findings presented in Fig. 3c solely depict the extension distance at specific moments and may not provide a complete representation of the complete intricacies of the liquid spreading behavior.

Generally, the work shows a fully phenomenological liquid directional transport, while the specific theoretical modeling remains an open topic. Here, it should be pointed out that the results obtained in this particular experimental context, involving continuous external liquid injection, exhibit differences when compared to pure self-driving mechanisms based on capillary forces. A notable example of a surface-tension-dominated self-driven bionic model is the Texas horned lizard's skin [54-56]. The presence of a structured skin enables liquids on its surface to spontaneously move in a specific direction. At this time, the asymmetric structure often presents a concave liquid interface in the spreading direction, which utilizes the Laplace pressure as the only driving force for liquid propulsion. However, in the experiments presented in this study, the spreading behavior is powered by the continual expansion of the fluid boundary, and the disparity in Laplace pressure resulting from the asymmetric structures dictates that the liquid consistently overcomes the pinning effect on one side, facilitating the final unidirectional transport. Consequently, the expanded liquid interface typically exhibits a convex shape on both sides, providing further evidence that the spreading behavior is not solely governed by surface tension but rather a composite effect. The injection flow rate is deliberately set to a small value to clearly observe the influence of surface tension on the directional spreading behavior. To provide an intuitive evaluation, the Weber number ($We$) for the directional transport experiment of the binary mixture (80% water + 20% ethanol) is calculated, i.e., $We = 16\rho v^2/(\pi^2 D^3 \gamma) = 0.013$, where $\rho$ (968 kg/m$^3$), $v$ (500 μL/min), $\gamma$ (38 mN/m), and $D$ (0.6 mm) are the liquid's density, flow rate, surface tension, and the diameter of syringe needle, respectively. At such a low Weber number, the inertia of the liquid can be considered negligible compared to its surface tension. In certain applications, it may be advantageous to employ a higher injection flow rate to achieve efficient transport. It will undoubtedly render the inertia of the fluid non-negligible, consequently weakening the influence of surface tension. This, in turn, may ultimately lead to the failure of directional transport, thereby resulting in the emergence of a dual-directional transport mode [8,10,14].

### 3.3. Functional liquid operations tailored by the surface

In comparison to the transportation of small-volume liquids on surfaces with micro/nano-scale features, the anti-gravity transport of large-volume liquids on millimeter-scale surfaces poses a challenging problem. This is primarily due to the substantial influence of body force (i.e., gravity, $\sim L^3$) compared to surface tension (which scales with $\sim L^1$) in the millimeter-scale regime, as dictated by scaling laws. Here, I demonstrate that the proposed surface design can leverage the magnetic field to enhance the performance of anti-gravity liquid transport.

Initially, the surface is placed on an adjustable inclined slope, presenting a standing status in the absence of the magnetic field. Upon injecting a liquid mixture (80% water + 20% ethanol) onto the surface, two distinct liquid transport situations can be observed. In situation 1, the liquid rises against gravity for a certain distance and then begins to transport downward; in situation 2, the liquid directly spreads downward. In the experiments, when the slope inclination is less than 18.2° (Fig. 4a(i)), the liquid first propagates upward due to a greater driving force ($F_a'$) than the downward component of liquid gravity along the slope ($G'$). With the increase of liquid volume, $G'$ surpasses $F_a'$, causing the downward transport of the liquid (referred to as situation 1). When the slope inclination surpasses 18.2° (Fig. 4a(ii)), the liquid becomes incapable of counteracting its own gravity ($F_a' < G'$), resulting in a unidirectional downward spreading (i.e., situation 2). In this context, I illustrate that the implementation of a magnetic field enhances the anti-gravity transport capability of the surface's pillars (when they are in an upward bending). As illustrated in Fig. 4b(i), the experimental results display the direct downward transport of liquids at an inclination of 18.2°. When the surface's pillars are actuated to bend upward, the surface shows an improved capability to counteract the force of gravity, thereby transitioning from situation 2 to situation 1. Specifically, as shown in Fig. 4b(ii), the liquid undergoes an upward transport within the initial 12.4 seconds, after which it ceases to propagate upward and begins moving downward (as observed in the snapshot at 18 seconds). Vice versa, when the pillars bend downward under the actuation of the magnetic field, the anti-gravity transport ability is weakened compared to the surface where pillars are in an upright state (Fig. S6).

The adjustability of spreading directions offers enhanced flexibility in regulating liquid behaviors. Through effective utilization of the magnetic field, I demonstrate a spontaneous modal shift in liquid transport, encompassing bidirectional transport as well as unidirectional transport (rightward/leftward). Fig. 4c(i-iv) illustrates the regions of the surface induced by the applied magnetic field, progressing from left to right, which consist of standing pillars (in gray), gradient (+) pillars (in green), uniformly bent pillars (in orange), and gradient (-) pillars (in blue). When the liquid mixture (80% water + 20% ethanol) is injected at the junction of standing pillars and gradient (+) pillars, an initial bidirectional spreading behavior is observed due to the nearly equal leftward and rightward pinning effect (Fig. 4c(i)). When the rightward propagating liquid encounters the bending pillars with greater curvature, the bidirectional liquid transport will evolve into rightward unidirectional liquid transport (Fig. 4c(ii)). This phenomenon arises because the unbalanced Laplace pressure effect, induced by magnetic-driven bending, surpasses the unbalanced Laplace pressure effect stemming from the inherent curvature structure of the pillars. Subsequently, upon reaching the gradient (-) region, the reduced curvature of the pillars halts the rightward liquid spreading and initiates a leftward spreading, thus presenting a reverse spreading behavior (Fig. 4c(iii)-(iv)). The experimental illustration is shown in Fig. 4c(v), where the liquid exhibits dual-directional spreading within the first 9.8 seconds, followed by a shift to unidirectional transport in the rightward direction until 39.8 seconds, and ultimately undergoes a change in spreading direction towards the left until 49 seconds.

Synchronous operation of multiple liquids holds significant appeal for advanced liquid manipulation on surfaces. One intriguing application is the use of liquid mixing to transform the surface into a liquid reactor. In this study, I illustrate that the introduction of a local magnetic field can enable efficient liquid mixing. As illustrated in Fig. 4d(i), the liquids (20% ethanol + 80% water) in two syringes are simultaneously injected onto the surface's right side (blue liquid, located at the standing pillar region) and left side (red liquid, located at the rightward bending pillar region). Experimental results, as shown in Fig. 4d(ii), reveal that the blue liquid is unidirectionally transported towards left (i.e., the surface's intrinsic transport direction), while the red liquid, influenced by the rightward bending pillars, spreads towards the right. At approximately 20.5 seconds, the two liquids successfully converge through their respective operations. All above experiments clearly demonstrate the potential of magnetized surfaces for multi-functional liquid operations, particularly highlighting their remarkable tunability.

### 3.4. Multimodal and 3-DOF solid transport

In addition to its liquid manipulation capabilities, the bioinspired magnetized surface also exhibits the ability to transport solid objects through dynamic coordination of its pillars. This functionality draws inspiration from metachronal coordination observed in cross-scale natural systems. For instance, biological cilia utilize metachronal motion to generate efficient propulsion [4], while millipedes employ metachronal coordination to facilitate crawling [57]. Consequently, numerous biomimetic systems have recently been developed to enable robotic crawling [36,49] and dynamic liquid/solid transportation [18,37,38,44] through the implementation of metachrony. Regarding solid transportation, bioinspired magnetic surfaces have been engineered with the abilities for 1-DOF unidirectional transport [16,45] and 2-DOF omnidirectional transport [44]. Differently, the study represents a pioneering advancement by introducing distinct transport modes for the proposed surface. These transport modes are intricately dependent on both the shape and arrangement of the transported objects. Additionally, a novel magnetic control method is proposed to enable the 3-DOF solid transport, comprising a combination of rotation and translation.

As illustrated in Fig. 5a(i), the metachronal motion, which is employed for solid transport, can be accomplished by translating the magnet (with the left pole as N and the right pole as S). This translation of the magnet initiates a sequential beating cycle of the affected pillars, resulting in a phase difference among neighboring pillars and giving rise to a typical metachronal wave. The metachronal wave propagates towards the left direction, facilitating the asynchronous transport of the solid cargo, which occurs through four distinct states (Fig. 5a(ii)). In the initial state I, the cargo's supporting pillars remain unaffected by the magnetic field, thereby causing the cargo to remain stationary. In the backward movement state II, the gradient bending of pillars on one side induces backward movement of the cargo. In the forward movement state III, the pillars sequentially rebound, applying propulsive forces ($F_p$) to the cargo and propelling it forward. In the final state IV, the cargo regains stability and undergoes a short-distance movement in the same direction as the magnet's translation. Due to the coordination of nearby pillars, the solid is transported over a relatively short distance within a single cycle, compared to the extensive translation distance of the magnet. Thus, this phenomenon is referred to as "asynchronous transport". Furthermore, I evaluate the multi-cycle asynchronous transport of solids with varying weights. Remarkably, after six transport cycles, the lighter solid cargo (cardboard: approximately 0.4 g) exhibits a greater displacement of around 20 mm, whereas the heavier solid cargo (cardboard + plasticine: approximately 3.4 g) achieved a displacement of about 14 mm (Fig. 5a(iii)-(iv),Video S4).

Curved-shape objects present a new transport mode, described as the synchronous type (Fig. 5b(i)). The transport mechanism of the synchronous type can also be divided into four states, but it does not primarily rely on propulsive forces generated through friction with pillars. As illustrated in Fig. 5b(ii), when the pillars near the cargo are actuated, the cargo transitions from the initial state I to the captured state II. The capture is completed by the gradient arrangement formed through metachronal coordination of the pillars. With the magnet's translation, the gradient arrangement region undergoes synchronous positional changes. In this process, the combined effect of the curvature object's gravity and the propulsive force exerted by the pillars enables the cargo to continuously follow the magnet (i.e., synchronous transport). As the experimental verification given in Fig. 5b(iii), the capsule closely follows the magnet from one end of the surface to the other within a single translation cycle. Interestingly, the same capsule can switch from the synchronous transport mode to an asynchronous mode by simply altering the orientation of the capsule from vertical to horizontal (Fig. 5b(iv), Video S5). These findings enrich our understanding of the solid transportation facilitated by the magnetized surface and hold potential for advancements in multi-solid operations, such as separation and classification.

Moreover, I develop a magnetic actuation strategy for achieving 3-DOF solid transportation, as shown in Fig. 6a(i). The specific strategy is to create an asymmetric distribution of propulsion forces on two sides of the axis (the red dashed line) by controlling different numbers of pillars. The resultant force that deviates from the center of gravity (denoted as point O) will generate an additional torque to rotate the solid object while transporting it. As illustrated in Fig. 6a(ii), the translated magnet, following a path at a slight angle to the object's body axis, can excite this imbalanced transportation process. The gray area A signifies the pillars that are about to apply elastic propulsion to the object, while the yellow area B encompasses the pillars that have already returned to their original state. The area A maintains an asymmetric distribution relative to the object's axis (red dashed line), consequently triggering the torque-driven rotational transport of the object. Throughout statuses 1 to 5, the object completes a combination of rotational and translational movement through the sequential activation of the pillars positioned beneath it. In addition, this driving strategy possesses a corrective effect that ensures alignment between the central axis of the object and the translation direction of the magnet. Consequently, once the correction process is finalized through several cycles, the object reverts to a state of translational transport, as illustrated in Fig. 6b(i) and Video S6. The data presented in Fig. 6b(ii) further substantiate this conclusion, as it illustrates a progressively stable rotation angle (where clockwise rotation angle is defined as a positive value). It is noted that the center position of the object continuously changes the coordinates on the $x$-axis and $y$-axis, proving the existence of directional transport. Using a similar strategy can also realize the counterclockwise rotation of solids, as shown in Fig. 6c(i). The transported solid exhibits similar transport characteristics, with the only distinction being the alteration in rotational direction (Fig. 6c(ii) and Video S7). Additionally, the intrinsic curvature structure of the pillar may introduce additional effects to its beating (such as self-twisting [49]), which may affect the 3-DOF solid transport. However, these effects are mitigated by using a relatively weak magnetic field, that is, the ferrite magnet is placed about 9 mm away from the surface's substrate for actuation. In the future, it is expected that more interesting results and new application potential will burst out under the combination of the proposed driving strategy and the special beating mode of heterostructure pillars. The proposed solid transport strategy, by coordinating the magnetic field and object arrangement, opens up a new idea for the advancement of versatile and diverse solid transport modes. The demonstrated 3-DOF transport mode in this study represents merely a single facet of the numerous possibilities that can be explored.

## 4. Conclusions

In this work, I present a bioinspired multi-asymmetric magnetized surface that can tailor liquid operations under static magnetic and achieve multimodal solid transport with dynamic magnetic fields. The proposed surface exhibits a unique combination of simplicity (absence of complex external devices), stability (elimination of dynamic magnetic field control), and flexibility (adjustable transport behaviors) in liquid operations. To a certain extent, it balances the trade-off between fixed structure surfaces and stimuli-responsive surfaces. Meanwhile, the introduction of a rotational DOF complements the conventional 2-DOF solid transport surfaces. The magnetic control strategy, as proposed, effectively enables the 3-DOF rotational transport of solids. These notable findings collectively signify advancements when compared to prior studies (Table S2). Specifically, for liquid manipulation, the surface, with curvature structure design, can directly transport liquids in a specific direction. By introducing the magnetic field effect, uniform/gradient bending can be acquired to further regulate liquid transport behaviors, e.g., enhancing the transport speed and changing transport direction. And attractively, the modulation of the magnetic field can also make the anti-gravity transport have tunable properties (e.g., the critical angle for situation transition), enable the liquid to spontaneously change spreading directions, and perform the multi-liquid operation (e.g., liquid mixing). Additionally, the naturally inspired dynamic metachronal is developed via this surface for multi-modal solid transport. For different shapes of solid objects, the surface can transport them with the asynchronous or synchronous type. With the changed object arrangement, the same solid object can switch between the two transport modes. Furthermore, the three-DOFs rotational transport of solids is realized by coordinating the magnetic field and object arrangement. The proposed bioinspired surface holds the ability of transporting both liquids and solids, which has great application potential in the fields of small-scale manipulation devices, soft actuators, and functional surfaces.

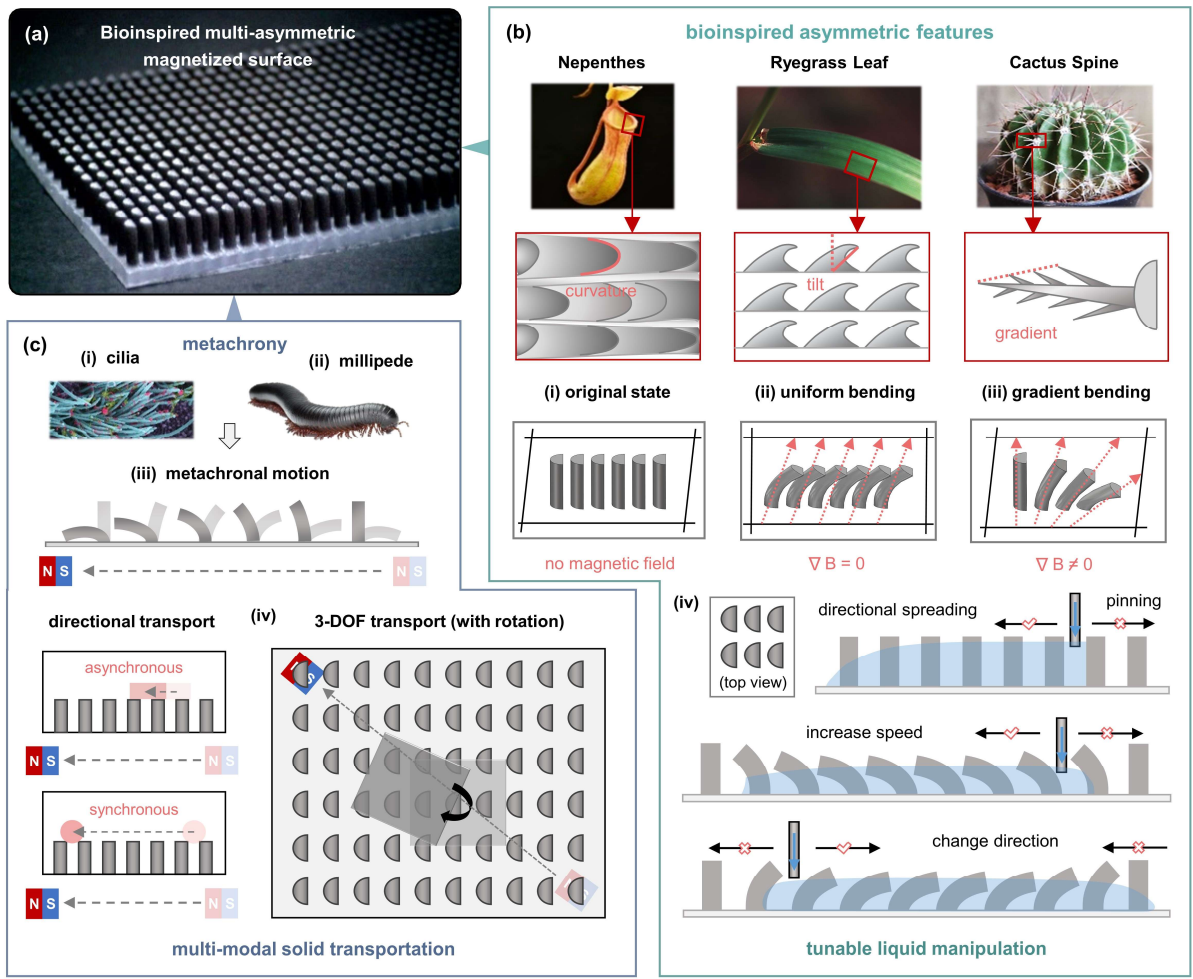

**Figure 1.** (a) Illustration of the bioinspired multi-asymmetric magnetized surface. (b) Schematic of the design, actuation principles, and functionalized liquid transport abilities of bioinspired asymmetric features: (i) curvature design of pillars on the surface (used for directional liquid transport); (ii)-(iii) magnetic-actuated tilt and gradient arrangement (used for increasing the liquid transport speed and changing the liquid transport direction); (iv) tunable liquid manipulation with three asymmetric features. (c) The metachrony in cross-scale biological systems (e.g., (i) cilia; (ii) millipede) inspired (iii)-(iv) multimodal and 3-DOF solid transportation.

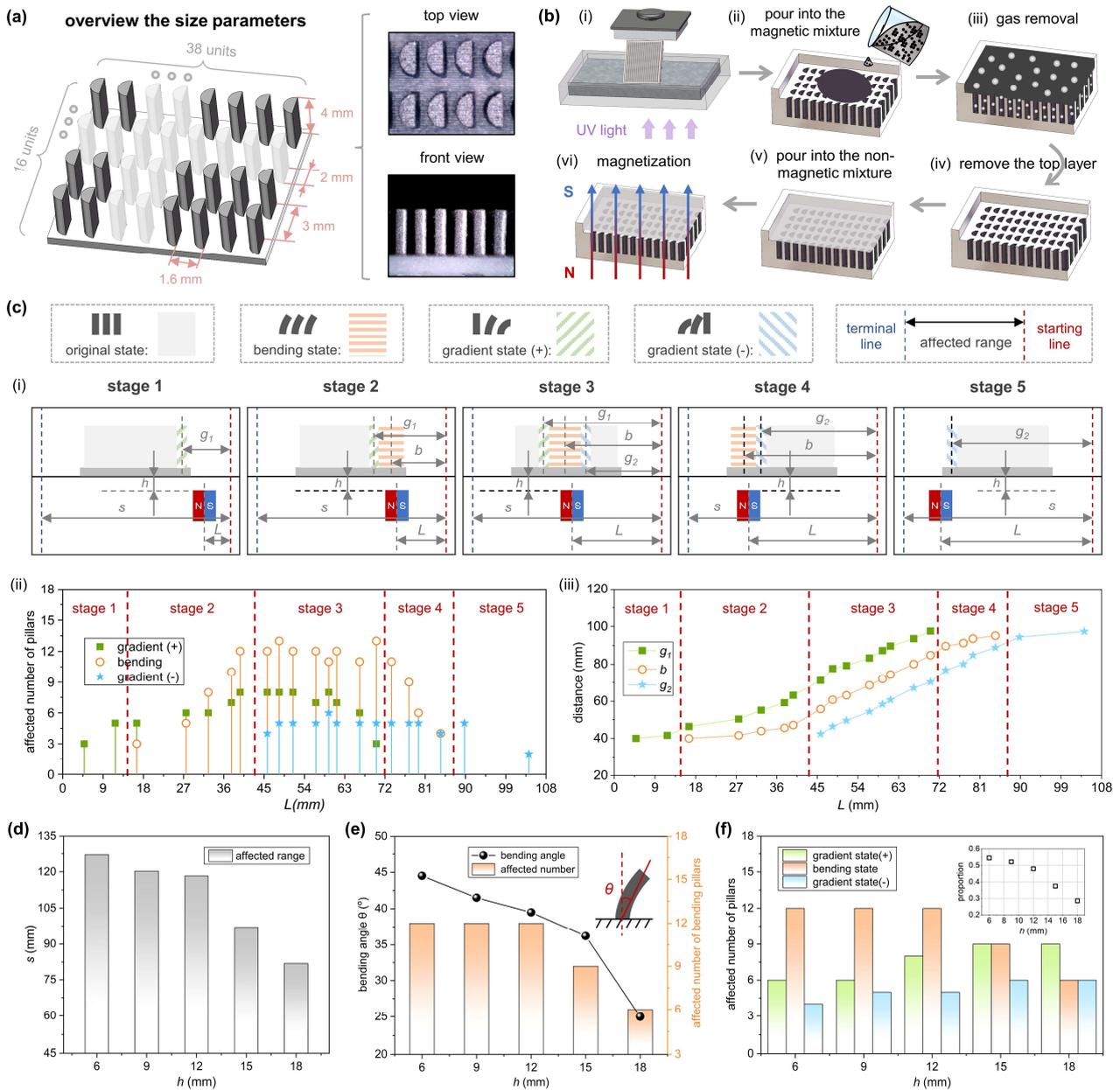

**Figure 2.** Fabrication and calibration of the magnetized surface. (a) Illustration of the surface's size parameters and the snapshots from the top view and front view. (b) Fabrication procedures (including six steps). (c) (i) As the magnet moves from right to left, there are five stages that partially/fully contain four main surface morphologies (i.e., original, uniform bending, gradient (+), and gradient (-) regions). (ii)-(iii) experimental records of the five stages and the variation (the number of affected pillars and the position) of different regions. (d)-(f) The effect of different magnetic fields (adjusted by changing $h$) on the surface's morphology.

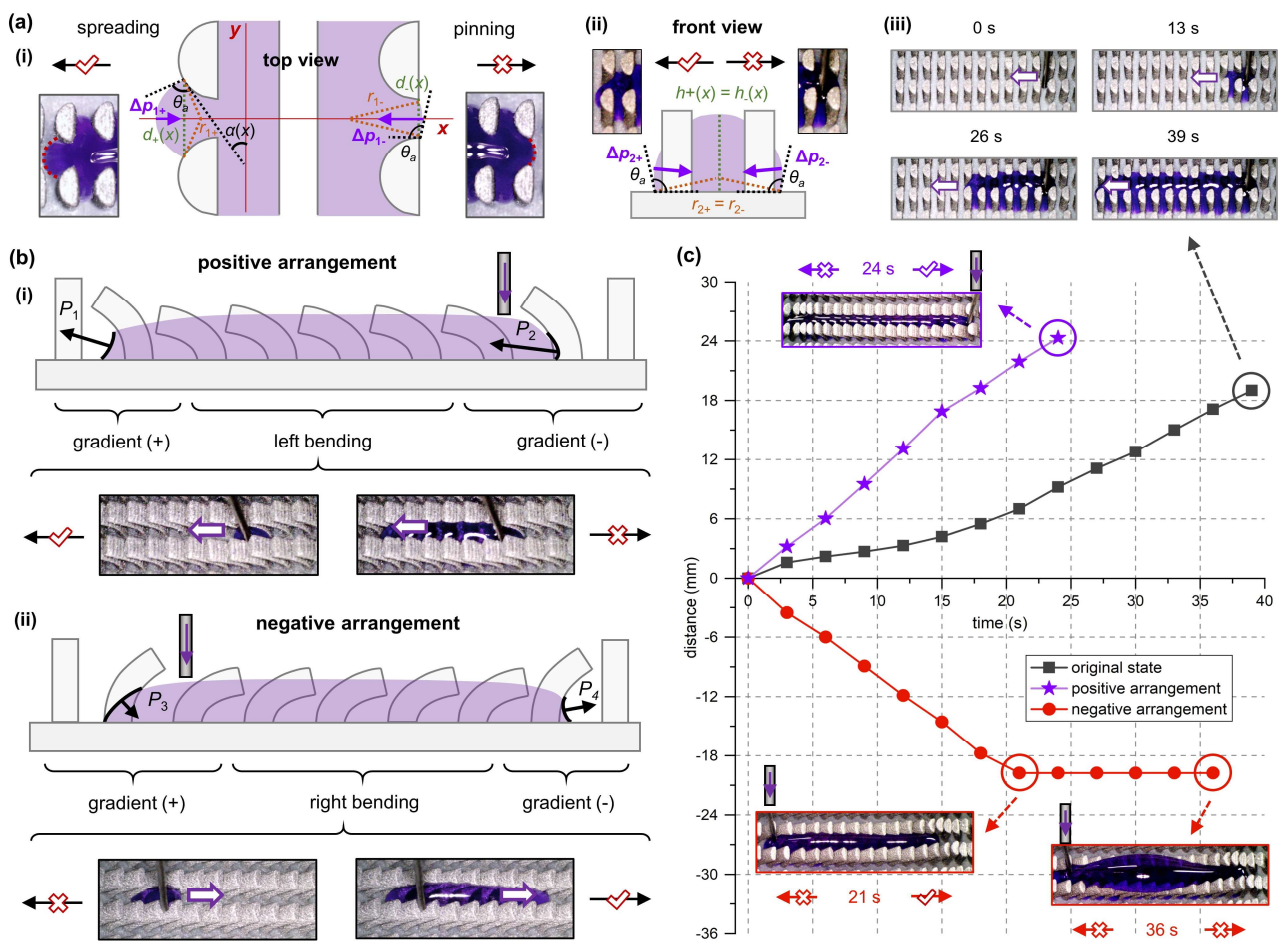

**Figure 3.** Mechanisms and experimental illustration of unidirectional liquid transport on the surface with different morphologies. (a) (i) Analysis of unbalanced Laplace pressures in the top view; (ii) analysis of unbalanced Laplace pressures in the front view; (iii) experimental snapshots of unidirectional liquid transport. (b) Mechanisms illustration and experimental snapshots of unidirectional liquid transport with (i) the positive arrangement, i.e., accelerating the transport speed, and (ii) the negative arrangement, i.e., changing the transport direction. (c) The time-distance curve of liquid directional spreading on the three types of arrangement.

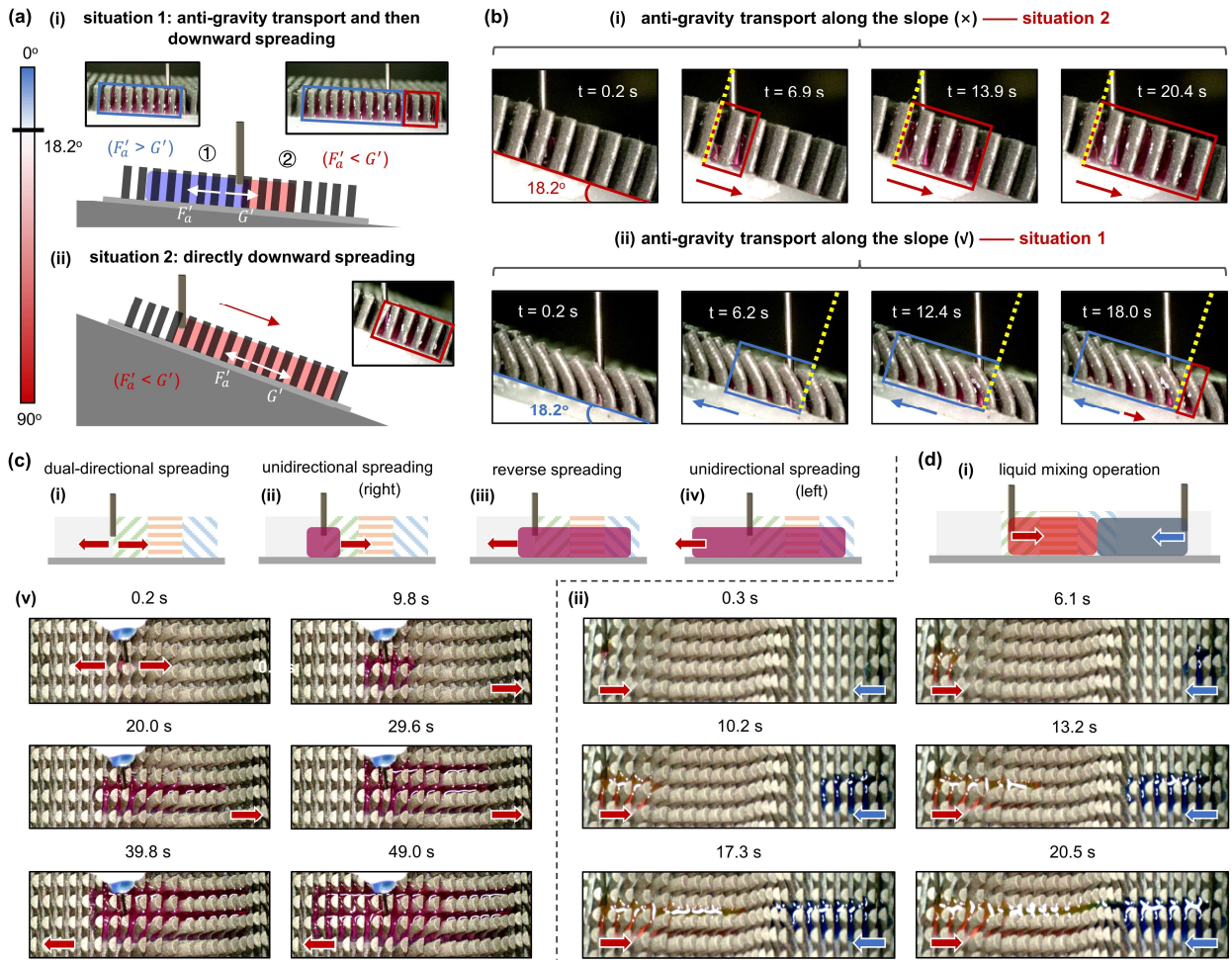

**Figure 4.** Multiple liquid operations via the magnetized surface. The adjustable anti-gravity climbing: (a) the liquid spreading on the surface placed on a slope contains two situations: (i) anti-gravity transport and then downward spreading; (ii) directly downward spreading. (b) The enhanced performance of anti-gravity transport under the effect of the magnetic field: (i) without the magnetic field; (ii) with the magnetic field. (c) The spontaneous modal shift in liquid transport: (i)-(iv) schematic diagram and (v) experimental illustrations. (d) The liquid mixing: (i) schematic diagram and (ii) experimental illustrations.

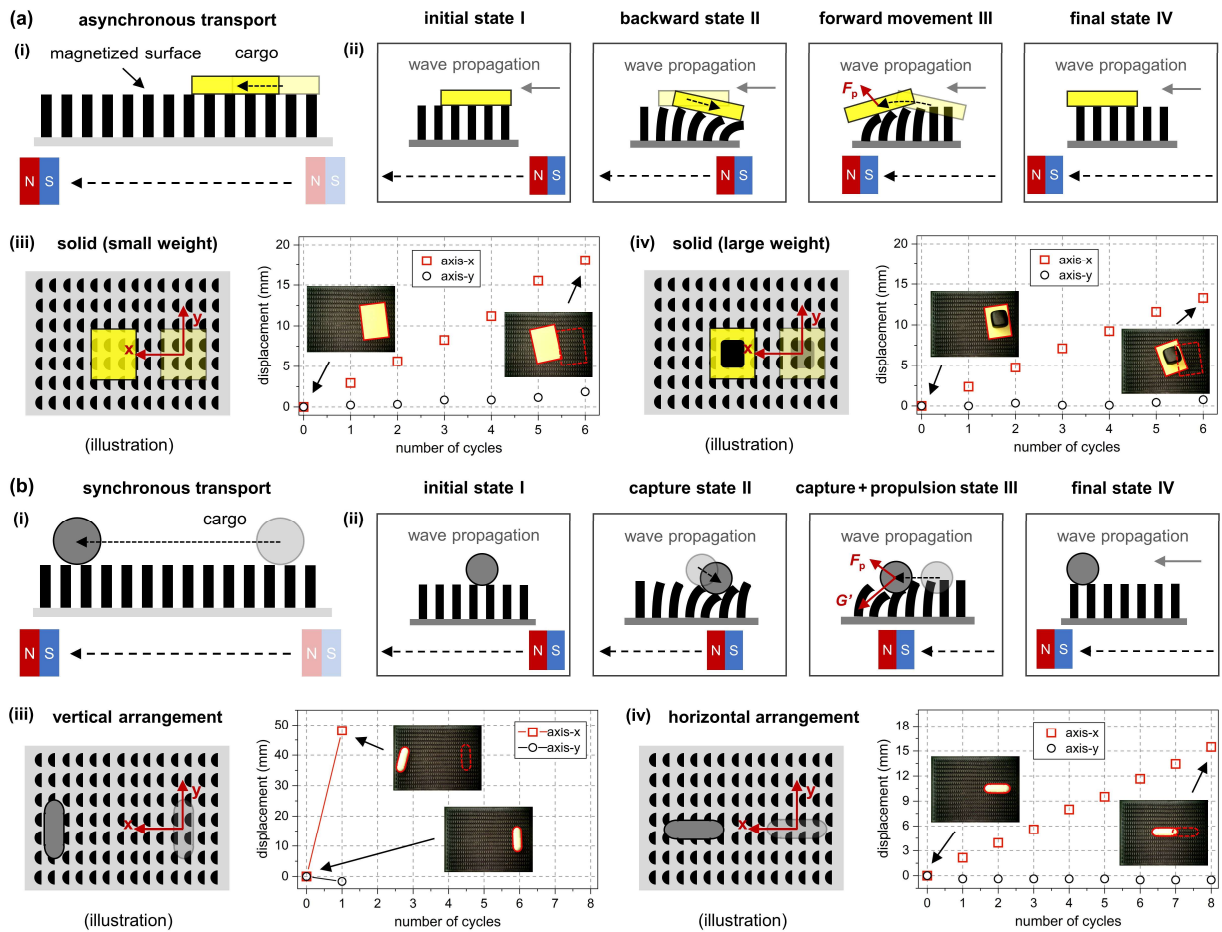

**Figure 5.** Illustration of the asynchronous and synchronous solid transportation. (a) Asynchronous transport mode. (i) schematic diagram; (ii) mechanism; and (iii)-(iv) experiments: the transport of the small-weight cargo and the large-weight cargo. (b) Synchronous transport mode. (i) schematic diagram; (ii) mechanism; and (iii) experiments: the synchronous transport of the capsule. (iv) when the capsule is rearranged from vertical to horizontal, the synchronous transport mode changes into the asynchronous mode.

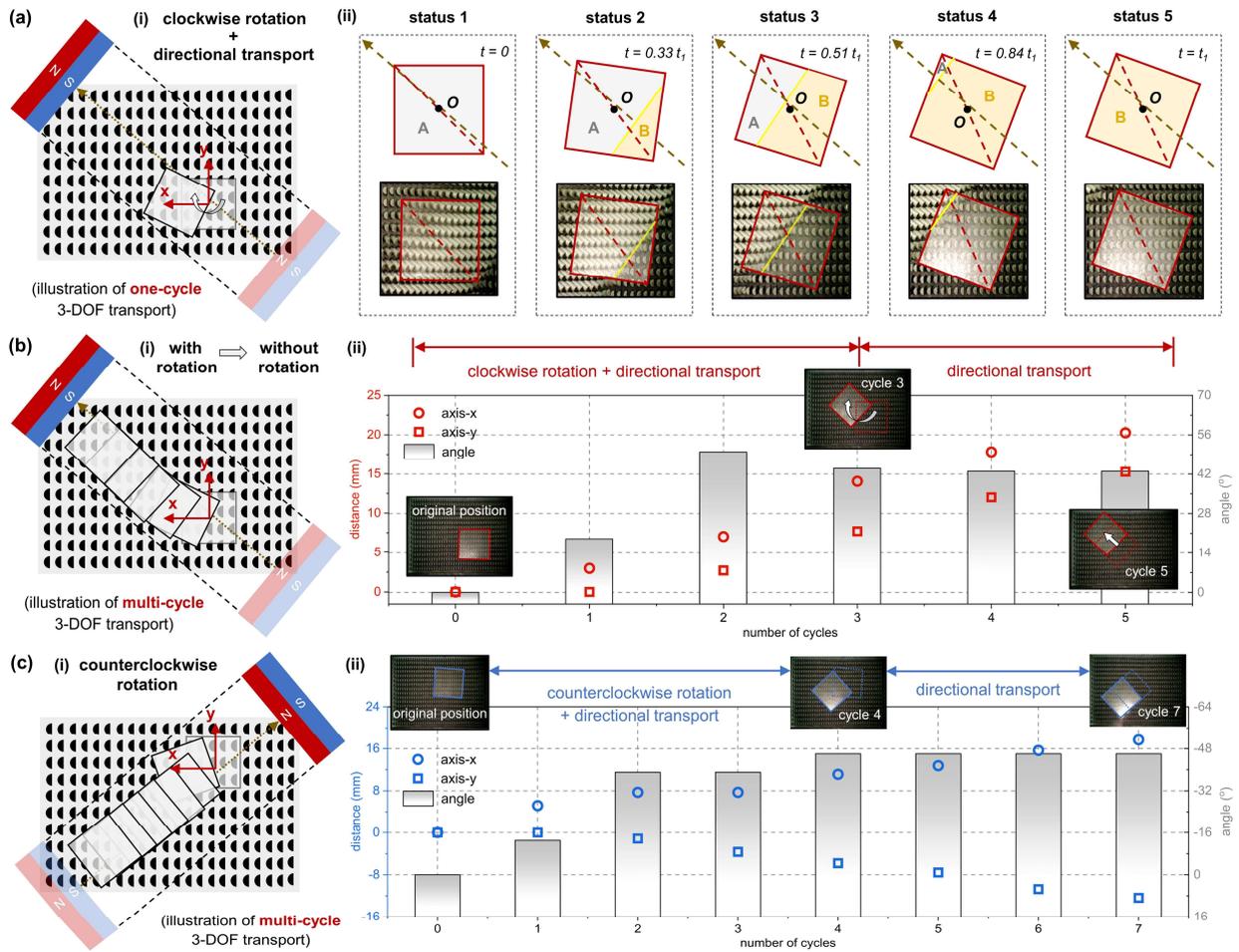

**Figure 6.** 3-DOF solid transportation via the magnetized surface. (a) (i) Illustration of one-cycle 3-DOF solid transport (clockwise type), (ii) including five statuses. (b) (i) Illustration of multi-cycle 3-DOF solid transport (clockwise type); (ii) within the first three cycles, it contains clockwise rotation and directional transport, then there is only directional transport in the last two cycles. (c) (i) Illustration of multi-cycle 3-DOF solid transport (counterclockwise type); (ii) within the first four cycles, it contains counterclockwise rotation and directional transport, then there is only directional transport in the last three cycles.